\documentclass[aps,prb,onecolumn,superscriptaddress,floatfix,showpacs,amsmath,amssymb,nofootinbib,longbibliography]{revtex4-2}% Physical Review

\usepackage{graphicx,xcolor}% Include figure files
\usepackage{array}[=2016-10-06] % MUST be before dcolumn/tabularx
\usepackage{dcolumn}
\newcolumntype{.}{D{.}{.}{-1}}
\usepackage{tabularx}

\usepackage{booktabs}
\usepackage{makecell}

\usepackage{bm}% bold math
\usepackage[utf8]{inputenc}
\usepackage{epstopdf}
\usepackage{subfigure}
\usepackage{amsmath,hyperref}
\usepackage{soul}
\usepackage{german}
\usepackage{nicefrac}
\captionsenglish
\usepackage{siunitx,xcolor}

\newcommand{\cpp}{$c_{\rm p}$}

\newcommand{\alvol}{$\alpha_{\rm v}$}

\newcommand{\ali}{$\alpha_{i}$}

\newcommand{\cpm}{$c_p^{\rm mag}$}
\newcommand{\cpb}{$c_p^{\rm bg}$}
\newcommand{\tn}{$T_{\rm N}$}

\newcommand{\lfs}{Li\(_2\)FeSiO\(_4\)}

\newcommand{\lzs}{Li\(_2\)ZnSiO\(_4\)}

\newcommand{\compound}{\lfs}%{Li$_2$FeSiO$_4$}

\newcommand{\wh}[1]{\textcolor{black}{#1}}
\newcommand{\rev}[1]{\textcolor{black}{#1}}

\newcolumntype{Y}{>{\centering\arraybackslash}X}
\newcolumntype{Z}{>{\centering\arraybackslash}p}

\begin{document}

%\preprint{APS/123-QED}
\title{Critical behavior and evidence of dimensional crossover in quasi-two-dimensional Li$_2$FeSiO$_4$}

%%%%  AUTHORS %%%%%%%%%%%%%%%%%%%%%%%%%%%%%%%%%%%%%%%%%%%%%%

\author{W.~Hergett}
\altaffiliation{Both authors contributed equally.}
\affiliation{Kirchhoff Institute of Physics, Heidelberg University, INF 227, D-69120 Heidelberg, Germany}

\author{\rev{K.~Ackermann}}
\altaffiliation{Both authors contributed equally.}
\affiliation{Institute for Theoretical Physics, Heidelberg University, Philosophenweg 19, D-69120 Heidelberg, Germany}

\author{\rev{E.~Walendy}}
\affiliation{Kirchhoff Institute of Physics, Heidelberg University, INF 227, D-69120 Heidelberg, Germany}

\author{S.~Spachmann}
\affiliation{Kirchhoff Institute of Physics, Heidelberg University, INF 227, D-69120 Heidelberg, Germany}

\author{M.~Jonak}
\affiliation{Kirchhoff Institute of Physics, Heidelberg University, INF 227, D-69120 Heidelberg, Germany}

\author{M.~Abdel-Hafiez}
\affiliation{Kirchhoff Institute of Physics, Heidelberg University, INF 227, D-69120 Heidelberg, Germany}
\affiliation{Physics Department, Faculty of Science, Fayoum University, Fayoum 63514, Egypt}
\affiliation{Department of Applied Physics and Astronomy, University of Sharjah, P. O. Box 27272 Sharjah, United Arab Emirates}

\author{\rev{M.~W.~Haverkort}}
\affiliation{Institute for Theoretical Physics, Heidelberg University, Philosophenweg 19, D-69120 Heidelberg, Germany}

\author{R.~Klingeler}
\email[Email:~]{klingeler@kip.uni-heidelberg.de}
\affiliation{Kirchhoff Institute of Physics, Heidelberg University, INF 227, D-69120 Heidelberg, Germany}

%%%%%%%%%%%%%%%%%%%%%%%%%%%%%%%%%%%%%%%%%%%%%%%%%%

\date{\today}% It is always \today, today,
             % but any date may be explicitly specified

\begin{abstract}
We report thermal expansion and heat capacity studies on Li$_2$FeSiO$_4$ single crystals which enable us to investigate the critical behavior in the magnetically quasi-two-dimensional (2D) material. Pronounced $\lambda$-shaped anomalies at the magnetic ordering temperature $T_{\rm N}$ imply significant magneto-elastic coupling.
\rev{Our analysis of both} the thermal expansion and the specific heat data implies the crossover from 2D Ising-like behavior for $|(T-T_{\rm N})/T_{\rm N}|>0.3$ to 3D Ising behavior \rev{below $\simeq 1.3\times T_{\rm N}$. The 2D-like behavior is further supported by density functional calculations which show minimal dispersion perpendicular to the crystallographic $ac$ planes of the layered structure, thereby indicating the 2D nature of magnetism at higher temperatures.} Our results extend the available model materials of quasi-2D magnetism to a high-spin $S=2$ system with tetrahedrally coordinated Fe$^{2+}$-ions, thereby illustrating how magnetic order evolves in a 2D Ising-like system with orbital degrees of freedom.

\end{abstract}

\maketitle

\section{Introduction}

The evolution of long-range magnetic order in two-dimensional magnetic materials sensitively depends on the nature of specific magnetic systems under study, with spin-orbit coupling being a key parameter~\cite{deJongh}. The Mermin-Wagner theorem rules out magnetic long-range order at finite temperatures in one- (1D) or two-dimensional (2D) Heisenberg systems with short-range interactions~\cite{MerminWagner} which was extended to the XY model and sufficiently decreasing long-range interactions~\cite{Bruno2001,Halperin2018}. In 2D XY magnets, Berenzinskii, Kosterlitz, and Thouless showed that only quasi-long-range order can develop at finite temperatures~\cite{Berezinsky1970,Kosterlitz1973,Kosterlitz1974}. While in 2D Ising systems, true long-range order occurs at a finite temperature~\cite{Onsager}. The presence of long-range magnetic order at $T>0$~K in magnetically anisotropic 2D van der Waals magnets highlights the relevance of magnetic anisotropy for the evolution of a magnetic ground state (see, e.g., Refs.~\onlinecite{Gong2019review,Wang2022} and references therein). Short-range interactions can be even large enough to allow the formation of magnetic order in finite 2D van der Waals magnets without any magnetic anisotropy~\cite{Jenkins2022}. The spin size, e.g., affects the relevance of quantum fluctuations, and effects of randomness may be relevant and drive a system into long-range order~\cite{deJongh,Crawford2013}.

Measurements of  thermodynamic response functions, namely the thermal expansion ($\alpha$) and the specific heat (\cpp ), provide sensitive tools to study the evolution of long- and short-range order. Analyzing the anomalies at the phase boundaries yields the critical exponent $\tilde{\alpha}$ describing\footnote{In this manuscript, the critical exponent is named $\tilde{\alpha}$ in order to discriminate the thermal expansion coefficient $\alpha$.} the critical behavior of \cpp\ or $\alpha \propto |t|^{-\tilde{\alpha}}$ ($|t|=|(T-T_{\rm N})/T_{\rm N}|$), thereby elucidating the qualitative nature of the system under study~\cite{Johansen1984,Scheer1992,Krellner2009,Spachmann2022,Kim2002,Kozlovskii2015,Zywocinski1997}. Dilatometric studies on single crystals in addition yield the uniaxial and hydrostatic pressure dependencies of ordering phenomena which in 2D materials is not only technologically relevant for tuning the materials by appropriate strain-engineering (see, e.g., ~\cite{Wang2020, Miao2021,Li2022}) but also allows to separate competing energy scales and to identify short-range ordering phenomena~\cite{Klingeler2005,Knafo2007,Wang2009,Takenaka2017,Gries2022,Reinold2024}. This also allows to study the quantum critical nature of phase transitions and has evidenced 2D quantum critical endpoints in layered Sr$_3$Ru$_2$O$_7$ and  Na$_2$Co$_2$TeO$_6$~\cite{Gegenwart2006,Arneth2025}. In van-der-Waals materials which in several examples have been shown to exhibit true long-range magnetic order down to the single-layer limit, dilatometric studies on bulk materials have been successfully applied to investigate the magnetic phase boundaries, the effect of uniaxial strain, and to decipher relevant microscopic parameters in $\alpha$-RuCl$_3$~\cite{He2018,Widmann2019,Schoenemann2020,Kocsis2022}, Cr$_2$Ge$_2$Te$_6$~\cite{Spachmann2022b,Spachmann2023}, CrI$_3$~\cite{Arneth2022} and CrCl$_3$~\cite{ali2025}.

\begin{figure}[tb]
	\includegraphics[width=0.75\columnwidth,clip] {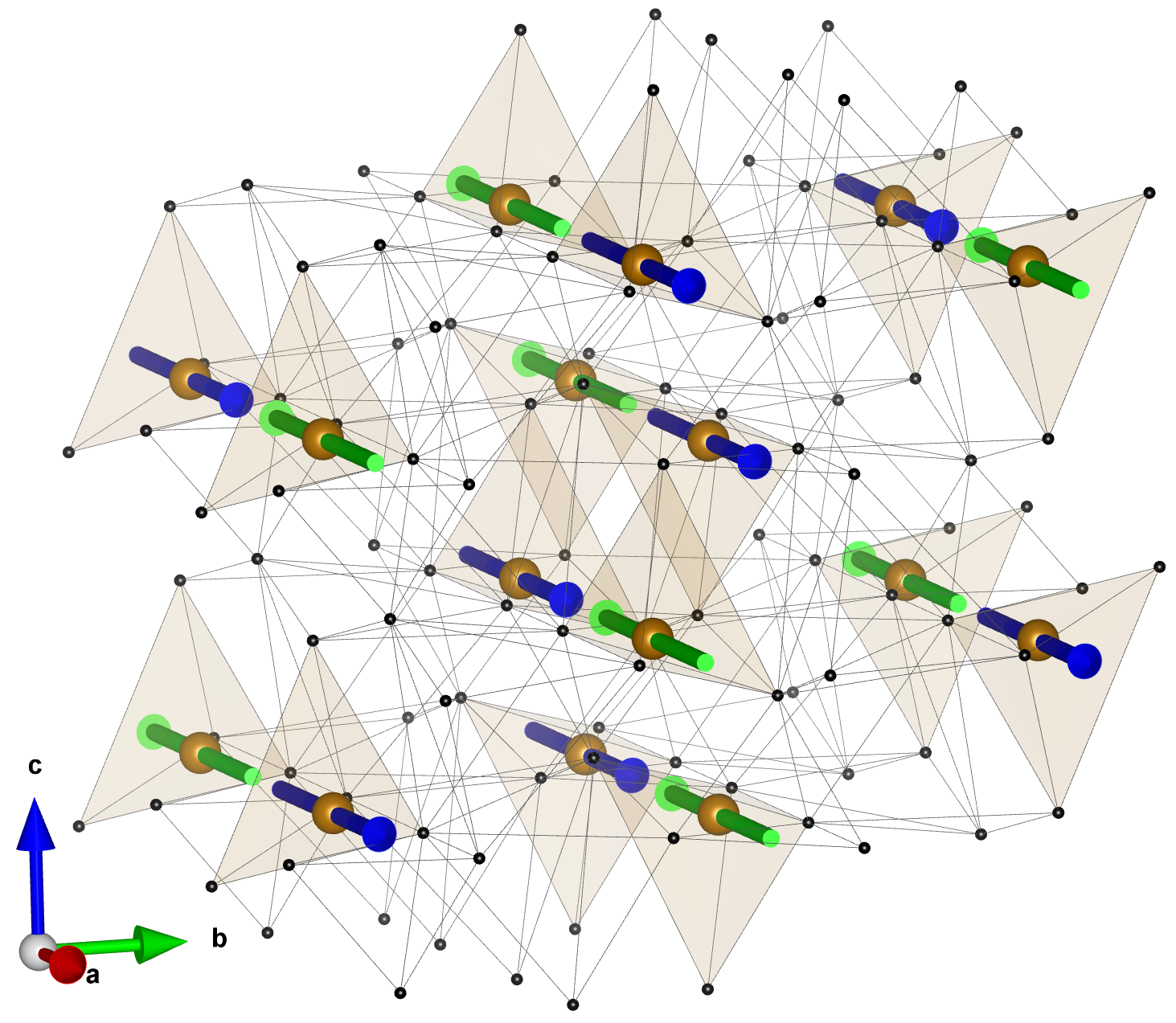}
	\caption{\wh{Schematics of the crystal and magnetic structure of \lfs . FeO$_4$ tetrahedra are shown in beige, with Fe$^{2+}$ ions located inside. Blue and green arrows indicate the orientations of the antiparallel magnetic moments. Small black spheres represent oxygen. The sketch was made with the VESTA software~\cite{Vesta} using structural data from CCDC 1859157~\cite{Hergett2019b}}.
 }\label{schematics}
\end{figure}

In Li\(_2\)FeSiO\(_4\) studied at hand, high-spin ($S=2$) Fe$^{\rm 2+}$ ions are arranged in a layered structure forming a magnetically quasi-2D system with dominating magnetic exchange coupling within the $ac$-layers while interlayer couplings are only weak and partly frustrated (see Fig.~\ref{schematics})~\cite{Lu2015,Nishimura2008,Hergett2025}. \wh{In the high-spin 3$d^6$ electronic configuration orbital degrees of freedom may be relevant which is confirmed by our numerical studies at hand.} Finite interlayer coupling and magnetic anisotropy yield long-range antiferromagnetic (AFM) order below $T_{\rm N}= 17$~K. Both a reduced ordered moment and the observed static hyperfine field indicate significant orbital contributions. Here, we report studies of the thermal expansion and of the specific heat on \lfs\ single crystals. We report the uniaxial and hydrostatic pressure dependencies and quantify magneto-elastic effects. Analysis of the anomalies at the phase transition shows that \lfs\ falls into the 3D Ising class. In addition, we find a distinct crossover to approximately 2D Ising behavior above \tn . This agrees to the 2D nature of magnetism in \lfs\ suggested by our numerical study and might be associated with orbital degrees of freedom.

\section{Experimental and numerical methods}

Millimeter-sized single crystals of the $Pmnb$-\lfs\ polymorph were grown by the high-pressure optical floating-zone method as described in detail in Refs.~\onlinecite{Hergett2019a,Hergett2019b,Hergett2021}. Oriented single crystals were cut into cuboids with approximate dimensions \(1.1 \times 1.1 \times 1.0~\text{mm}^3\) with respect to the crystallographic directions. High-resolution dilatometry measurements were performed using a three-terminal high-resolution capacitance dilatometer from Kuechler Innovative Measurement Technology~\cite{kuechler2012}, operated inside a variable-temperature insert in an Oxford magnet system~\cite{Werner2017}. The capacitance readout was facilitated by Andeen-Hagerling’s AH 2550A ultraprecision 1-kHz capacitance bridge. With the dilatometer, the uniaxial thermal expansion $dL_i(T)/L_i$ ($i=a,b,c$) and the linear thermal expansion coefficients $\alpha_i = 1/L_i \times dL_i(T)/dT$ along the crystallographic axes were studied at temperatures ranging from 2 to 300~K in zero field.

\rev{The electronic structure was investigated using density functional theory as implemented in the FPLO~\cite{FPLO1999} code, which included either scalar-relativistic corrections for the Schr\"{o}dinger equation or solving of the full relativistic four-component Dirac equation. The crystal structure was obtained from Ref.~\cite{Hergett2019b}. The calculations employed the Perdew-Wang 92 (PW92) Local-density approximation (LDA) functional~\cite{PerdewWang1992} assuming a paramagnetic state, and the Brillouin zone was sampled with a $40\times20\times40$ k-point mesh. Otherwise, the default settings were used. To confirm the LDA results, the calculations were repeated using the Perdew–Burke–Ernzerhof (PBE) Generalized Gradient approximation (GGA) exchange-correlation functional~\cite{PBE-PhysRevLett.77.3865}. The downfolding to maximally-projected atomic-like Wannier functions~\cite{FPLOWAN2003} was performed using the inbuilt functionality of FPLO.}

\section{Experimental results}

\subsection{Anomalies in thermal expansion and heat capacity at $T_{\rm N}$ }

%\begin{figure}[bt]
%	\includegraphics[width=1.0\columnwidth,clip] {JTdistorsion.png}
%	\caption{\wh{Electronic level scheme for Fe$^{2+}$ in tetrahedral coordination, showing the qualitative energy shifts of the $d$-orbitals of the ideal ($T_d$) structure, as a result of the Jahn--Teller instability.}
    %https://doi.org/10.1007/s00269-011-0468-6
% }\label{JT}
%\end{figure}

%\begin{figure}[tb]
 %   \centering
    %\includegraphics[width=1.0\columnwidth, clip]{Fig1-DLL.png}
    %\caption{Relative length changes, $dL_i(T)/L_i$ ($i = a, b, c$), of \lfs. The solid black line denotes the volume expansion \dllvol~=~$1/3\times \sum_i{dL_i/L_i}$; the vertical dashed line indicates \tn .
%}
    %\label{dll}
%\end{figure}

\begin{figure}[tb]
    \centering
    \includegraphics[width=0.6\columnwidth, clip]{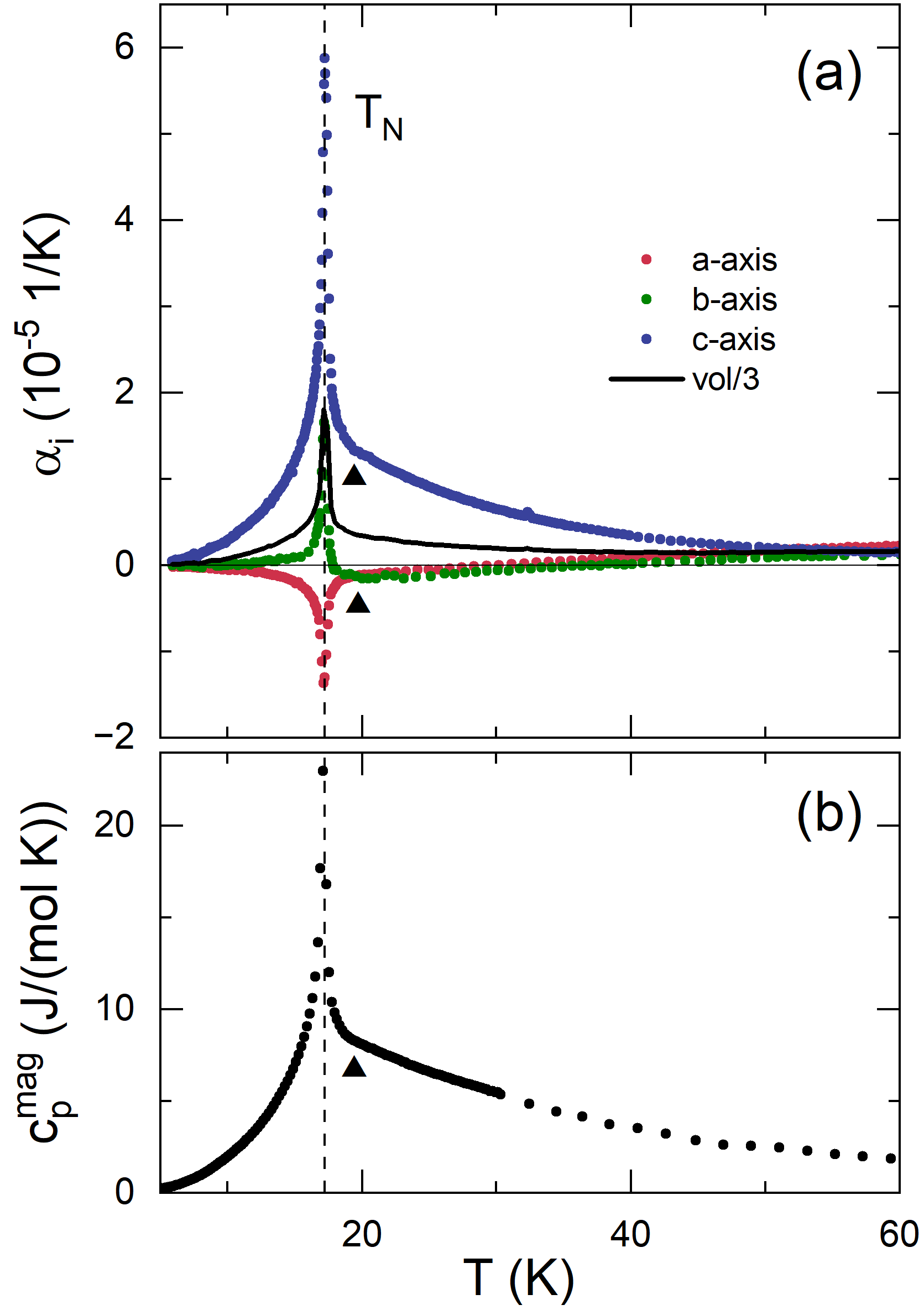}
    \caption{(a) Linear and volume thermal expansion coefficients $\alpha_i$ ($i = a,b,c, \rm{vol}$) along the three crystallographic axes and for the volume (solid black line). (b) Magnetic specific heat \cpm\ obtained by correcting the measured specific heat by the lattice contribution from measurements of the non-magnetic analog material \lzs~\cite{Hergett2025}. The vertical dashed line indicates \tn. Black triangles mark the dimensional crossover region as discussed in Sec.~\ref{ch:critical}.}
    \label{alpha_cp}
\end{figure}

\rev{The thermal expansion coefficients \ali\ ($i=a,b,c$ axis and the volume) of \lfs\ presented in Fig.~\ref{alpha_cp} are dominated by sharp anomalies at the long-range magnetic ordering temperature $T_{\rm N} = 17.2(5)$~K. Their $\lambda$-like character implies the continuous nature of the phase transition.} This is resembled by the anomaly in the magnetic specific heat, \cpm, which has been obtained as the difference between the measured specific heat of \lfs\ and a background \cpb\ given by its non-magnetic analog \lzs~\cite{Hergett2025} (Fig.~\ref{alpha_cp}b). The contrasting signs and magnitudes of the thermal expansion anomalies along the different crystallographic axes, at \tn , reveal significant and anisotropic magneto-elastic coupling, manifested in a contraction of the $b$ and $c$ axes and an expansion of the $a$ axis at \tn. The volume features a negative anomaly, too, which implies that hydrostatic pressure will yield the suppression of \tn . In addition to the pronounced anomalies at \tn , there is a wide temperature regime well above \tn\ of anisotropic temperature dependence of the uniaxial thermal expansion coefficients. Concomitantly, the specific heat data evidence non-phononic entropy changes in this temperature regime which we attribute to short-range
magnetic order~\cite{Hergett2025}.

\subsection{Analysis of critical behavior\label{ch:critical}}

\begin{figure}[tb]
    \centering
\includegraphics[width=0.8\columnwidth, clip]{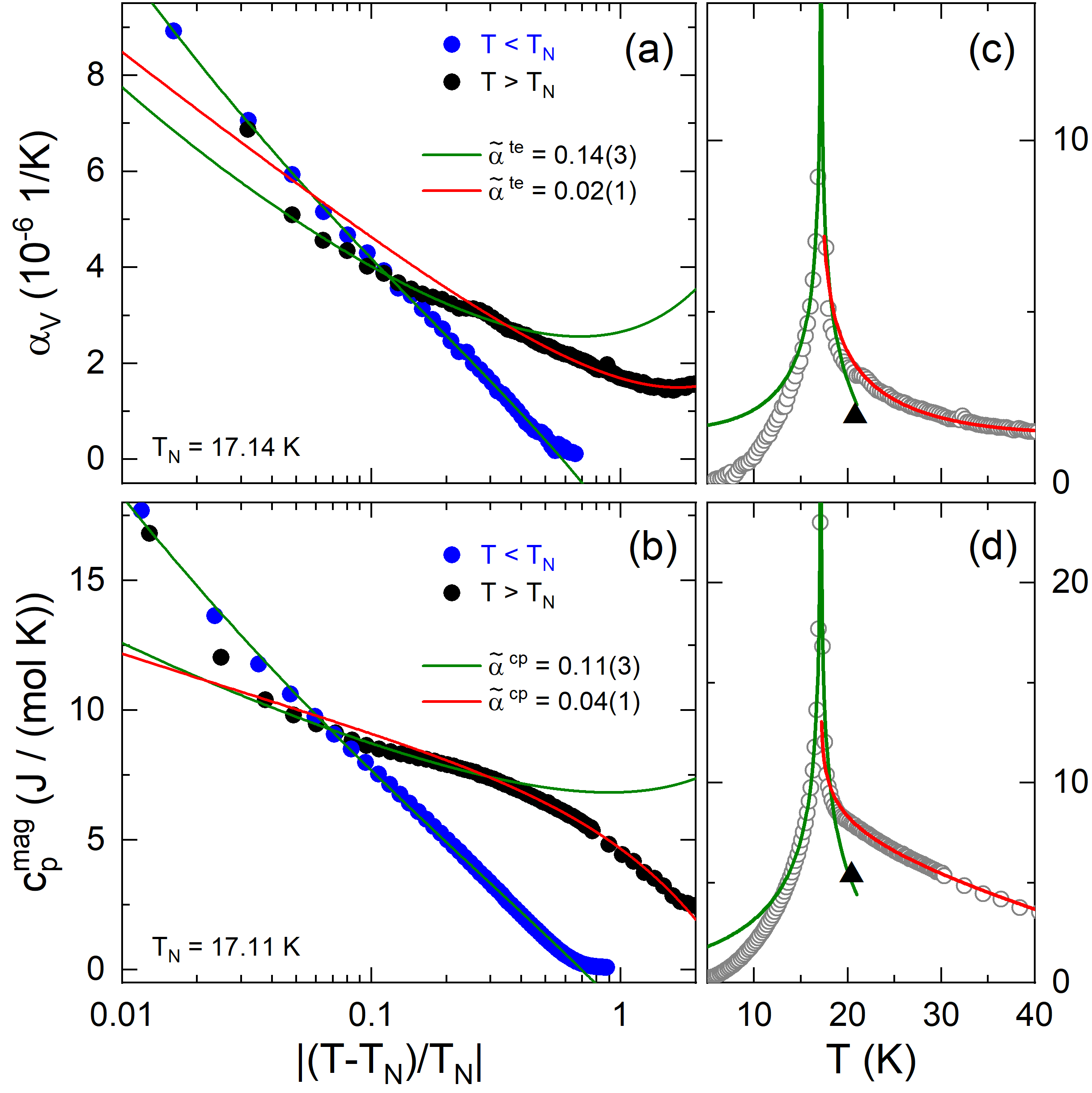}
\caption{(a,b) Semi-logarithmic plots of the volume thermal expansion coefficient \alvol\ and the magnetic specific heat \cpm\ vs. the reduced temperature $\left|t\right| = \left| (T - T_{\rm N})/T_{\rm N} \right|$ for \tn~=~17.14~K and \tn~=~17.11~K, respectively. Blue dots (black dots) mark data for $T < T_{\rm N}$ ($T > T_{\rm N}$). Lines are fits to the data using Eq.~\eqref{eq:critical} for different values of the critical exponent $\tilde{\alpha}^{\rm cp,te}$ (see  text). (c,d) Experimental data and obtained fit functions. Triangles mark the crossover region.}
    \label{fig3}
\end{figure}

Analysis of the critical behavior at continuous phase transitions provides key information on the universal nature of the systems and the ordering phenomena under study. Measurements of the thermodynamic response functions, i.e., the thermal expansion and the specific heat, are sensitive tools to study the evolution of long- and short-range magnetic order in quasi-2D magnets as they reflect the critical behavior of the system's energy fluctuations. Among the thermodynamic quantities that exhibit singularities at phase transitions, the isobaric heat capacity, $c_{\rm p}(T)$, is the most commonly studied. The singularity in $c_{\rm p}(T)$, characterized by the critical exponent $\tilde{\alpha}$ (see footnote 1), is typically described by the following equation~\cite{Pelissetto2002,Kornblit1978,Scheer1992}:

\begin{equation}
c_{\rm p}(t) = \frac{A}{\tilde{\alpha}} |t|^{-\tilde{\alpha}} + B + E t,
\label{eq:critical}
\end{equation}

for $T > T_{\rm N}$, where $t = (T - T_{\rm N})/T_{\rm N}$, and similarly for $T < T_{\rm N}$, using the same functional form with primed quantities (\cpp$'$, \alvol$'$, $T'_{\rm N}$) and parameters ($A'$, $\tilde{\alpha}'$, $B'$, $E'$). The first term represents the critical behavior while the background contribution is captured by $B+Et$. Both the magnetic specific heat \cpm\ and the volume thermal expansion coefficient \alvol\ presented in Fig.~\ref{alpha_cp} are evaluated using Eq.~\ref{eq:critical}, as shown in Fig.~\ref{fig3}. The constraints $E=E'$ and $T_{\rm N} = T'_{\rm N}$ were imposed without any significant loss in the quality of the fit. Furthermore, when data points around \tn\ were excluded from the analysis, a common best-fit parameter $\tilde{\alpha}=\tilde{\alpha}'$ could also be applied. Evaluation of the thermal-expansion (te) data yields $\tilde{\alpha}^{\rm te} = 0.14(3)$ for the fit in the temperature range 8.3~K~$<T<$~16.9~K and 18~K~$<T<$~21~K. In the range 22.4~K~$<T<$~83.2~K, $\tilde{\alpha}^{\rm te} = 0.02(1)$ is obtained. Similarly, analysis of the magnetic specific heat \cpm\ yields $\tilde{\alpha}^{\rm cp} = 0.11(3)$ for the fit in the temperature range 4.9~K~$<T<$~16.9~K  and  17.8~K~$<T<$~20.8~K. For 21.2~K~$<T<$~48.9~K, $\tilde{\alpha}^{\rm cp} = 0.04(1)$ is obtained. Table~\ref{tab:alpha_exp} summarizes the results of the critical analyses.

\begin{table}[tb]
    \centering
\caption{Critical exponents $\tilde{\alpha}$ obtained by fitting Eq.~\protect\ref{eq:critical} over different temperature ranges, along with an overview of theory predictions of $\tilde{\alpha}$ for various model systems.}
    \begin{tabularx}{0.95\columnwidth}{p{2.5cm}  Z{1.7cm}  Y}
%    \toprule
\hline \hline
Analysis of: &         $\tilde{\alpha}$ & Fit range [K]\\
%\midrule
\hline
Thermal  exp.        &  0.14(3) & 8.3-16.9 \& 18-21 \\
coefficient \alvol\  &  0.02(1) & 22.4-83.2 \\
\hline
%specific  &         $\alpha^{\rm cp}$ & Fit Range\\
Specific &          0.11(3) & 4.9-16.9 \& 17.8-20.8 \\
heat \cpp\  &         0.04(1) & 21.2-48.9 \\      %\hline
\multicolumn{3}{c}{~ }\\

%\hline \hline
Model &         $\tilde{\alpha}$ & Ref.\\
\midrule
         3D Ising  & 0.11 &\cite{Campostrini19993DIsing} \\
         2D Ising  & 0 & \cite{Onsager}\\
         3D Heisenberg  & -0.134 &  \cite{Pelissetto2002} \\
         Mean field  & 0 & \\
         3D XY  & -0.014 & \cite{Campostrini20013DXY} \\
        \hline \hline
%\bottomrule
\end{tabularx}
    \label{tab:alpha_exp}
\end{table}

% \begin{table}[tb]
%     \centering
% \caption{Critical exponents $\tilde{\alpha}$ obtained by fitting Eq.~\protect\ref{eq:critical} over different temperature ranges, along with an overview of theory predictions of $\tilde{\alpha}$ for various model systems.}
%     \begin{tabular}{l|c|c}
% %\hline \hline
% Analysis of: &         $\tilde{\alpha}$ & Fit range [K]\\
% \hline
% Thermal  exp.&  ~0.14(3)~ & 8.3-16.9 \& 18-21 \\
% coefficient \alvol\  &       0.02(1) & 22.4-83.2 \\
% \hline
% %specific  &         $\alpha^{\rm cp}$ & Fit Range\\
% Specific &          0.11(3) & 4.9-16.9 \& 17.8-20.8 \\
% heat \cpp\  &         0.04(1) & 21.2-48.9 \\      %\hline
% \multicolumn{3}{c}{~ }\\

%    % \hline %\hline
% Model &         $\tilde{\alpha}$ & Ref.\\
% \hline
%          3D Ising  & 0.11 &\cite{Campostrini19993DIsing} \\
%          2D Ising  & 0 & \cite{Onsager}\\
%          3D Heisenberg  & -0.134 &  \cite{Pelissetto2002} \\
%          Mean field  & 0 & \\
%          3D XY  & -0.014 & \cite{Campostrini20013DXY} \\
%     %        \hline \hline
% \end{tabular}
%     \label{tab:alpha_exp}
% \end{table}

Consistently, around \tn , our analysis yields critical exponents $\tilde{\alpha}^{\rm cp,te}$ of 0.11-0.14. Variations in the choice of \tn\ lead to slightly larger critical exponent values, ranging from 0.2 to 0.25, while variation of the fitting range affects the results only to a minor extent (see \cite{Cook1973} for a discussion of the role of constraints). For temperatures above 22~K, the extracted critical exponents decrease significantly to near-zero values (see Table~\ref{tab:alpha_exp}). The unambiguously positive value of $\tilde{\alpha}$ in the vicinity of \tn\ clearly excludes Heisenberg or XY behavior~\cite{Pelissetto2002}. It also questions the 2D Ising scenario suggested by the analysis of the order parameter from M\"{o}ssbauer ($\beta = 0.116$) and neutron ($\beta = 0.185$)  studies~\cite{Hergett2025}. In contrast, our findings are in agreement to the predictions of the 3D Ising model~\cite{Campostrini19993DIsing,Pelissetto2002}. The fact that, near to \tn , both the measurements of the specific heat and the thermal expansion are described by the 3D Ising scenario supports the validity of our analysis.

In both quantities, the behavior above $\sim 22$~K implies a different temperature dependence of \alvol\ and \cpp\ as described by nearly vanishing values of $\tilde{\alpha}$. Our findings hence evidence a change in the critical behavior from nearly vanishing $\tilde{\alpha}$ at high temperatures to $\tilde{\alpha}\simeq 0.11-0.14$ in the vicinity of \tn\ (see Table~\ref{tab:alpha_exp}). We note that the associated changes in the temperature dependencies of the thermal expansion coefficients and in the specific heat are  clearly visible in the experimental data presented (see the black filled triangles in Fig.~\ref{alpha_cp}). The divergence of both \cpp\ and \alvol , around \tn , is consistent with a 3D Ising behavior. In contrast, very small critical exponents observed at higher temperatures are within the range expected for the 2D Ising model~\cite{Onsager}.~\footnote{While $\tilde{\alpha}\simeq 0$ in general agrees with the mean-field prediction, mean-field behavior is excluded by the large temperature regime of short-range order extending far above \tn .} Our data hence suggest a dimensional crossover from 2D to 3D Ising behavior. This change in dimensionality appears above \tn\ within the short-range ordered regime reported in Ref.~\cite{Hergett2025}.

\rev{\section{Numerical studies}}

\begin{figure}[htb]
    \centering
    \includegraphics[width=0.8\columnwidth, clip]{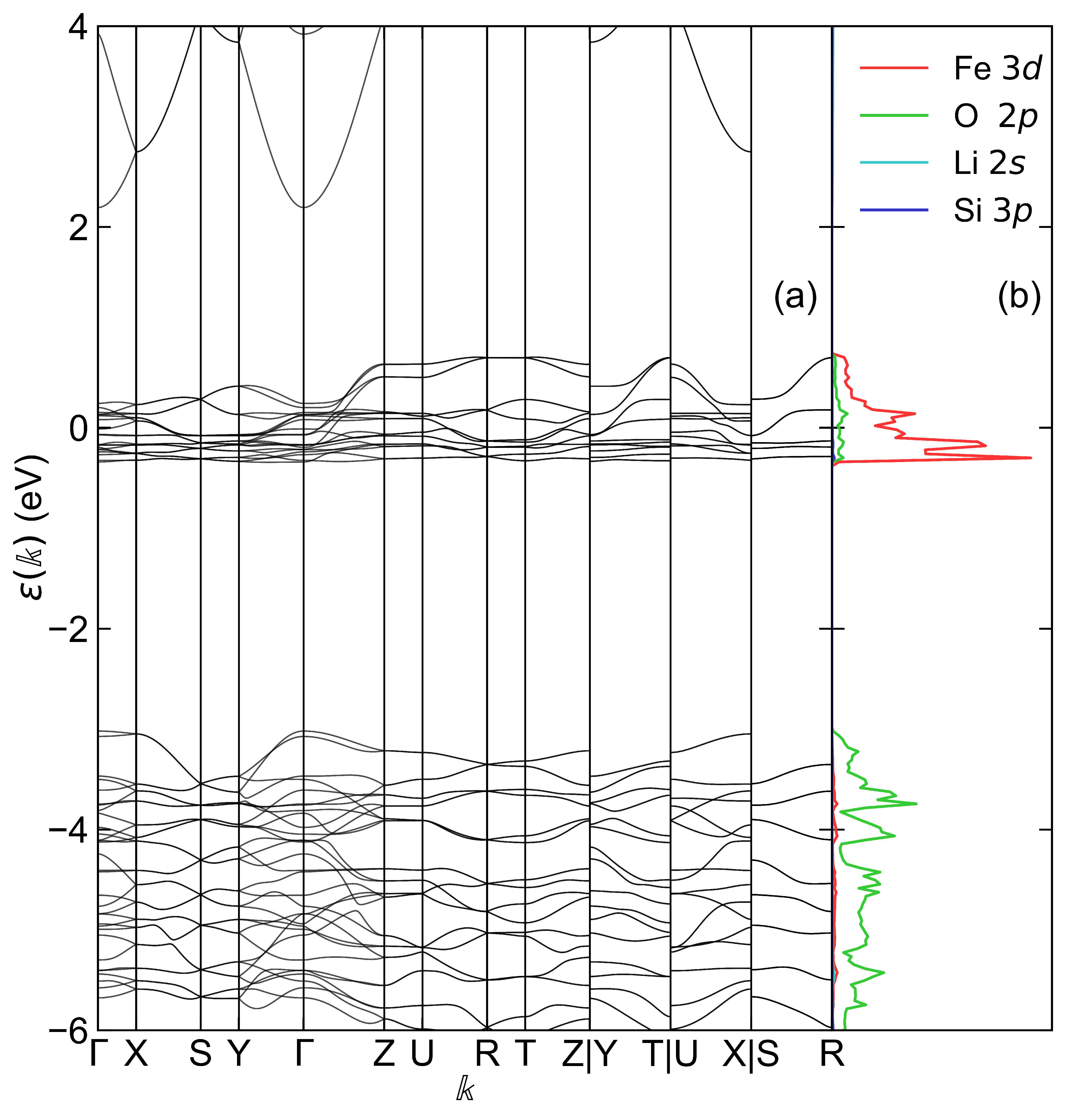}
    \caption{\rev{(a) Paramagnetic band structure of Li$_2$FeSiO$_4$ as computed with FPLO \cite{FPLO1999} using the PW92 functional. (b) Projected density of states for the Fe $3d$, O 2$p$, Li 2$s$ and Si 3$p$ orbitals.}}
    \label{fig:bandLDA}
\end{figure}

\begin{figure}[tb]
    \centering
    \includegraphics[width=0.5\columnwidth, clip]{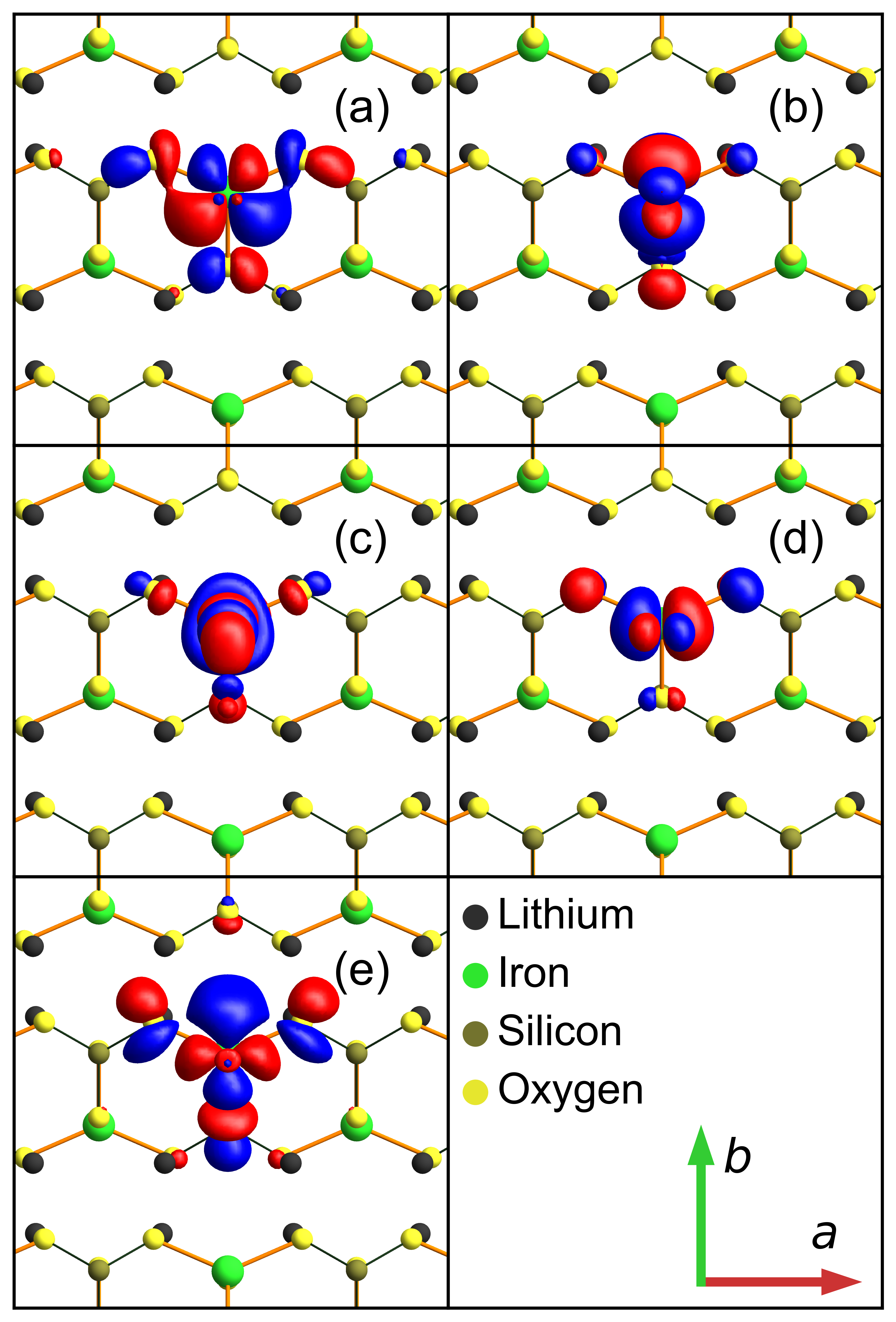}
    \caption{\rev{Isosurface plots of the Wannier functions localized on the iron  shown in cut through the crystal structure parallel to the $ab$-plane. The Isosurface is chosen such that 90\% of the electronic density is contained within. The subplots correspond to the (a) $3d_{{xy}}$, (b) ${3d}_{{yz}}$, (c) ${3d}_{{z}^2}$, (d) ${3d}_{{xz}}$, and (e) ${3d}_{{x}^2-{y}^2}$-like Wannier functions.}}
    \label{fig:wanPlot}
\end{figure}

\rev{The numerically obtained electronic band structure (Fig.~\ref{fig:bandLDA}a) reveals a set of isolated bands near the Fermi level which are primarily of iron $3d$ orbital character (Fig.~\ref{fig:bandLDA}b). These bands are rather flat and energetically well separated by a gap of approximately 3~eV from the occupied oxygen 2$p$ bands below and by 2~eV from the unoccupied bands above. Their isolation in energy around the Fermi energy strongly suggests that the magnetism in this system is driven by these localized iron $3d$ electrons.
%\rk{@Kevin: If I remember our discussion correctly, the following is true: The orbital degeneracy of the $3d$ levels is fully lifted with the ${3d}_{{x}^2-{y}^2}$-like Wannier function being the energetically lowest one, followed by ${3d}_{{xz}}$, $3d_{{xy}}$, ${3d}_{{yz}}$, and  ${3d}_{{z}^2}$. This means that in the ground state we have the ${3d}_{{x}^2-{y}^2}$ state doubly occupied and all other singly.}
To further investigate the electronic structure of \compound, the isolated iron $3d$ bands were downfolded to atomic-like Wannier functions~\cite{FPLOWAN2003} centered around the iron sites (see Fig.~\ref{fig:wanPlot}). The Wannier functions are highly confined within their respective layers. The only exception is the $3d_{x^2-y^2}$-like Wannier function displayed in Fig.~\ref{fig:wanPlot}e which has a small tail crossing into the next layer. This spatial confinement, indicating that hybridization occurs almost exclusively within the $ac$ planes of the layered structure, provides strong evidence for the quasi-two-dimensional nature of the magnetic interactions in \compound. This quasi-2D character is further supported by the band dispersion itself. Along momentum-space paths perpendicular to the layers ($\mathrm{\Gamma}\rightarrow\mathrm{X}$, $\mathrm{S}\rightarrow\mathrm{Y}$, $\mathrm{Z}\rightarrow\mathrm{U}$ and $\mathrm{R}\rightarrow\mathrm{T}$), the iron $3d$ bands are nearly flat. This minimal dispersion is in contrast to the dispersion exhibited along in-plane directions, reinforcing, together with the shape of the Wannier functions, the conclusion of weak inter-layer magnetic coupling.}

\begin{figure}[htb]
    \centering
    \includegraphics[width=0.5\columnwidth, clip]{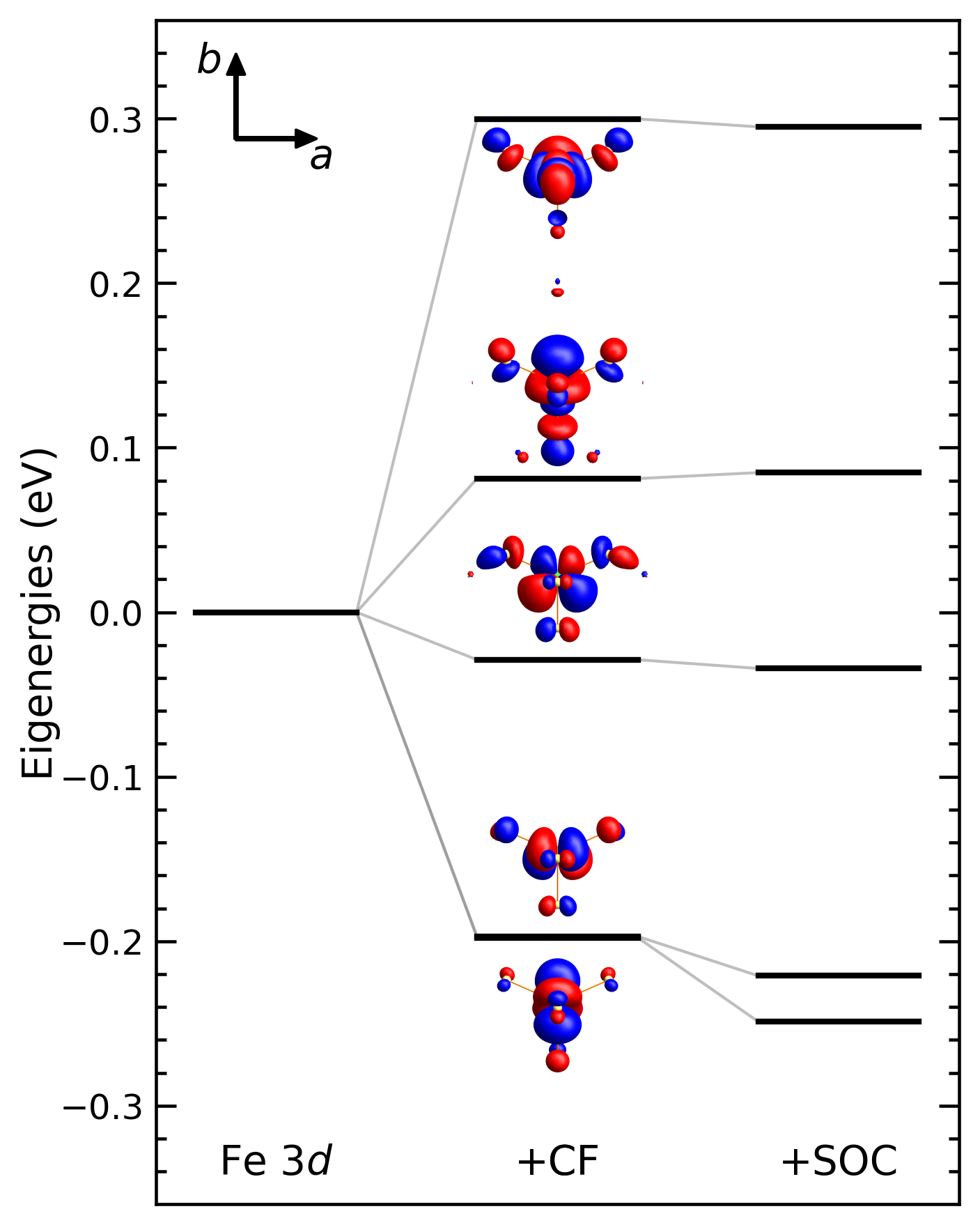}
    \caption{%The energy level diagram for the $3d$-like Wannier functions localized on iron derived by diagonalizing the local Wannier Tight-Binding Hamiltonian. The left section of the diagram presents an illustrative depiction of the degenerate $3d$ orbitals of the free ion. The central section features the scalar-relativistic LDA Hamiltonian, which accounts for the crystal field effects. Furthermore, the right section incorporates the full relativistic LDA Hamiltonian, which includes the effects of spin-orbit coupling. Additionally, in the central column, the crystal field eigenfunctions are displayed adjacent to their corresponding energy levels. The isosurfaces depicted are configured to encompass 90\% of the electronic density within each orbital.
    \rev{The energy level diagram (for PW92) for the $3d$-like Wannier functions localized on iron derived by diagonalizing the local Wannier Tight-Binding Hamiltonian. In comparison to the degenerated $3d$ orbitals of the free ion (left), the solution of the scalar-relativistic LDA Hamiltonian, which accounts for the crystal field effects (middle), and the effect of incorporating the full relativistic LDA Hamiltonian, which includes the effects of spin-orbit coupling, are displayed. Additionally, the crystal field eigenfunctions are displayed adjacent to their corresponding energy levels. The isosurfaces depicted are configured to encompass 90\% of the electronic density within each orbital.}
    }
    \label{fig:energyLevel}
\end{figure}

\rev{The process of downfolding to localized Wannier functions yields a tight-binding representation of the targeted electronic bands. By constraining this Hamiltonian to encompass only interactions within a local iron $3d$ shell, one derives the local crystal-field Hamiltonian. Diagonalization then yields the local eigenenergies (Table \ref{tab:ap:localEigenEnergies}) and eigenfunctions, which are illustrated in Fig.~\ref{fig:energyLevel}. At the scalar relativistic level, the two lowest eigenfunctions exhibit an approximate degeneracy, with an energy separation of $\Delta E_{\rm{CF}}^{\rm{PW92}} = 0.8$~meV. This degeneracy is lifted by including spin-orbit coupling in the DFT calculations by solving the full relativistic Dirac equation, which increases the energy gap to $\Delta E_{\rm{CF+SOC}}^{\rm{PW92}} = 27.9$~meV \wh{($\approx 324$~K)}.}

\rev{To investigate whether the small crystal-field gap is physical, analogous calculations were conducted utilizing the PBE functional. The resulting band structure is shown in the Appendix in Fig.~\ref{fig:ap:bandLDA-PBE} \rev{and the associated energy level diagram in Fig.~\ref{fig:ap:energyLevel-PBE}}. The relevant features for the quasi-two-dimensional nature of magnetism in \compound, specifically the isolated nature of the iron $3d$ bands and their flatness along certain paths through the Brillouin zone, remain largely unchanged compared to PW92. However, the crystal-field gap is increased to $\Delta E_{\rm{CF}}^{\rm{PBE}} = 10.4$~meV and the gap in the full relativistic case subsequently becomes $\Delta E_{\rm{CF+SOC}}^{\rm{PBE}} = 32.9$~meV \wh{($\approx 382$~K) (see Fig.~\ref{fig:ap:energyLevel-PBE} in the Appendix)}. The density functional calculations therefore suggest that the orbital degeneracy is completely lifted by the combined effect of the crystal-field and spin-orbit coupling.}\\

\section{Discussion and summary}

\rev{The quasi-2D nature of magnetism in \lfs\ which had been previously found in experimental magnetic susceptibility, M\"{o}ssbauer and neutron studies is confirmed by the obtained electronic band structure which implies highly confined Wannier functions with the exception of the Fe $3d_{x^2-y^2}$-like orbital. It in particular shows that magnetism is associated with energetically well isolated Fe $3d$ electrons. Hybridization occurs almost completely in the $ac$ planes of the crystallographic structure and hence confirms the quasi-2D nature of magnetic interactions. Our results show that the combined effects of the crystal field and of spin-orbit coupling completely lift orbital degeneracy of the Fe $3d$ levels so that \lfs\ provides a natural framework for the observed crossover from 2D to 3D behavior.}

Crossover in the critical behavior of energy fluctuations is widely studied by means of ultracold gases (see~\cite{Bloch2008} and references therein) but has been only rarely reported in solid state materials. One example is EuTe where $\tilde{\alpha}$, obtained from thermal expansion and specific heat data, assumes sizable finite values only when approaching \tn~\cite{Scheer1992}. In Ga$_{1-x}$Mn$_x$As, the critical exponent obtained from thermal diffusivity data shows a crossover from 3D Ising-like to 3D mean-field-like behavior~\cite{Yuldashev2012}. In this respect, we also note the potential effect of orbital degrees of freedom. It was shown for perovskite manganites that the Jahn-Teller effect can modify the system’s dimensionality and interaction range, leading to a non-standard universality class~\cite{Abdulvagidov2020}. For example, critical behavior consistent with the 3D Ising model has been observed La$_{1-x}$Sr$_{1+x}$MnO$_3$ around optimal doping, at the ferromagnetic transition~\cite{Lin2000,Kim2002}. In \lfs , there is no indication of a cooperative Jahn-Teller transition in the investigated temperature regime up to 400~K which is in agreement with complete lifting of orbital degeneracy as predicted by our numerical studies. However, orbital degrees of freedom may be relevant well above \tn .

In summary, our detailed studies of thermal expansion and heat capacity on Li$_2$FeSiO$_4$ single crystals enable us to investigate in detail critical behavior in the magnetically quasi-2D material. We find a dimensional crossover from approximately 2D Ising-like behavior for $|(T-T_{\rm N})/T_{\rm N}|>0.3$ to 3D Ising behavior which is evidenced by our analysis of the critical behavior of both the thermal expansion and the specific heat. \rev{Our analysis of the DFT band structure and the shape of the down folded Wannier functions yields minimal dispersion perpendicular to the $ac$ planes, indicating the 2D nature of the magnetic interactions.} Our data extend available model materials of quasi-2D magnetism to a high-spin $S=2$ system with tetrahedrally coordinated Fe$^{2+}$-ions, thereby illustrating how long-range magnetic order evolves in a 2D Ising-like system.

%\textit{critical analysis following the method of Kornblit and Ahlers: Magnetization, susceptibility, and specific heat measurements were made on a single crystal of La\(_{0.75}\)Sr\(_{0.25}\)MnO\(_3\). The ferromagnetic-to-paramagnetic phase transition was found at 346 K and four critical exponents were measured as: \( \alpha = 0.05 \pm 0.07 \), \( \beta = 0.40 \pm 0.02 \), \( \gamma = 1.27 \pm 0.06 \), and \( \delta = 4.12 \pm 0.33 \). The values of critical exponents are all between mean-field values and three-dimensional (3D) Ising-model values. The scaling behavior is well obeyed for all measurements, and the associated exponent relations are well satisfied, validating the critical analysis.}~\cite{Kim2002}

\begin{acknowledgments}
Valuable discussions with Manfred Salmhofer and Stefan-Ludwig Drechsler as well as experimental support from Christoph Neef are gratefully acknowledged. We also acknowledge the support by Deutsche  Forschungsgemeinschaft (DFG) under Germany’s Excellence Strategy EXC2181/1-390900948 (The Heidelberg STRUCTURES Excellence Cluster).\\

\end{acknowledgments}

%\hfill
%\newpage

\appendix
\section*{Appendix}\label{Sec_Appendix}

\renewcommand{\thefigure}{A\arabic{figure}}
\setcounter{figure}{0}

\renewcommand{\thetable}{A\arabic{table}}
\setcounter{table}{0}

\begin{figure}[htb]
    \centering
    \includegraphics[width=0.5\columnwidth, clip]{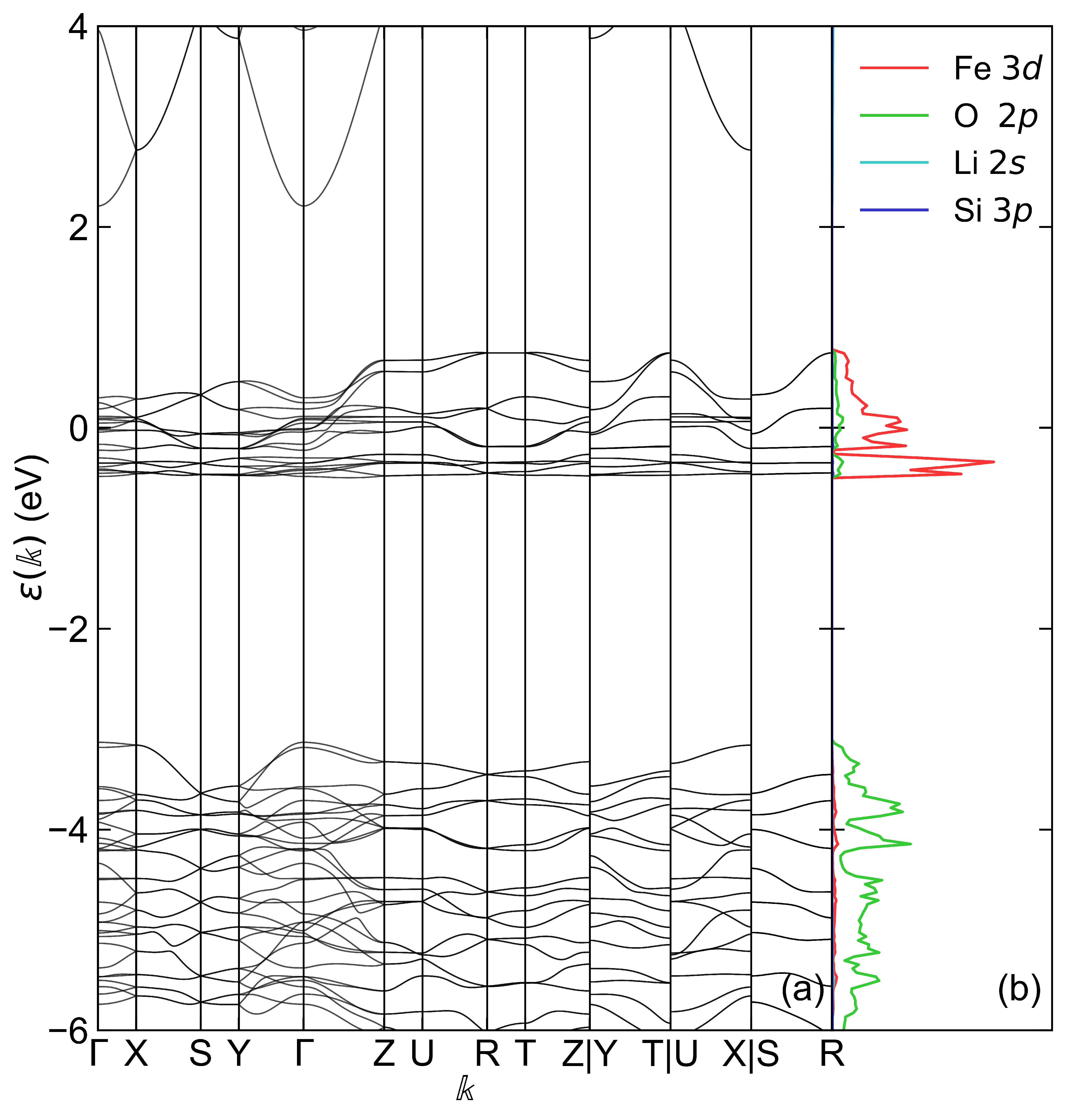}
    \caption{\rev{(a) Paramagnetic band structure of Li$_2$FeSiO$_4$ as computed with FPLO \cite{FPLO1999} using the PBE functional. (b) Projected density of states for the Fe $3d$, O 2$p$, Li 2$s$ and Si 3$p$ orbitals.}}
    \label{fig:ap:bandLDA-PBE}
\end{figure}

\begin{figure}[htb]
    \centering
    \includegraphics[width=0.5\columnwidth, clip]{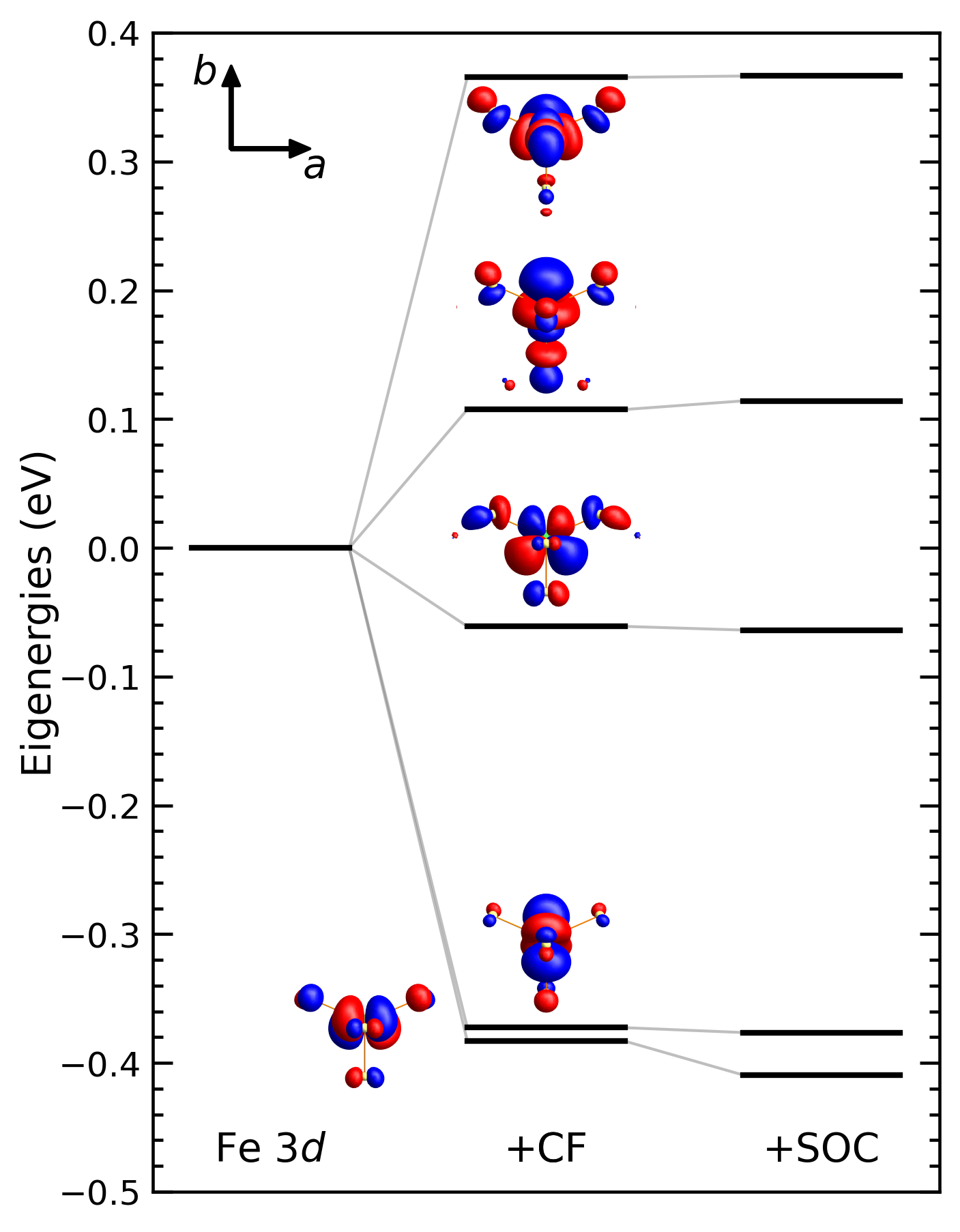}
    \caption{\rev{The energy level diagram (for PBE) for the $3d$-like Wannier functions localized on iron derived by diagonalizing the local Wannier Tight-Binding Hamiltonian. In comparison to the degenerated $3d$ orbitals of the free ion (left), the solution of the scalar-relativistic GGA Hamiltonian, which accounts for the crystal field effects (middle), and the effect of incorporating the full relativistic GGA Hamiltonian, which includes the effects of spin-orbit coupling, are displayed. Additionally, the crystal field eigenfunctions are displayed adjacent to their corresponding energy levels. The isosurfaces depicted are configured to encompass 90\% of the electronic density within each orbital.}}
    \label{fig:ap:energyLevel-PBE}
\end{figure}

\rev{The Appendix provides the local eigenenergies obtained by diagonalization of the local crystal-field Hamiltonian (Table \ref{tab:ap:localEigenEnergies}) as well as the band structure (Fig.~\ref{fig:ap:bandLDA-PBE}) and the associated energy level diagram (Fig.~\ref{fig:ap:energyLevel-PBE}) obtained using the PBE functional. }

\begin{table}[htb]
    \centering
    \caption{\rev{The eigenenergies ($E_{1-5}$) associated with the local Wannier Hamiltonian are computed utilizing the PW92 and PBE functional frameworks. Additionally, these computations are carried out with scalar relativistic corrections and fully relativistic. }}

    \begin{tabularx}{0.95\columnwidth}{Z{2cm} *{4}{Y}}
        \toprule
        %Eigenstate & $E_{\rm{CF}}^{\rm{PW92}}$ & $E_{\rm{CF}}^{\rm{PBE}}$ & $E_{\rm{CF+SOC}}^{\rm{PW92}}$ & $E_{\rm{CF+SOC}}^{\rm{PBE}}$\\
        \makecell{Relativistic \\ treatment} & \multicolumn{2}{c}{\makecell{Scalar relativistic \\ corrections}} & \multicolumn{2}{c}{\makecell{Full relativistic \\ treatment}} \\

        \cmidrule(lr){2-3} \cmidrule(lr){4-5}

        Functional & PW92 & PBE & PW92 & PBE \\
        \midrule
        $E_1$~(meV) & $-198$ & $-383$ & $-249$ & $-409$ \\
        $E_2$~(meV) & $-197$ & $-373$ & $-221$ & $-376$ \\
        $E_3$~(meV) & $\:\:-29$ & $\:\:-61$ & $\:\:-34$ & $\:\:-64$ \\
        $E_4$~(meV) & $\:\:\;\;\,81$ & $\;\;\,107$ & $\:\:\;\;\,85$ & $\;\;\,114$ \\
        $E_5$~(meV) & $\;\;\,300$ & $\;\;\,365$ & $\;\;\,295$ & $\;\;\,366$ \\
        \bottomrule
    \end{tabularx}

    \label{tab:ap:localEigenEnergies}
\end{table}

\bibliography{Grueneisen_biblio}

\end{document}